\documentstyle[12pt]{article}
\def\baselinestretch{1.5}

\begin{document}
\thispagestyle{empty}
\null\vskip -1cm
\centerline{
\vbox{
\hbox{November 24, 1997}\vskip -9pt 
\hbox{hep-ph/xxxxxxx}\vskip -9pt
     } \hfill
} \vskip 1cm
\centerline
{\large \bf
Interplay Between Strong and Weak CP Phases}
\vspace{1cm}
\centerline           {
David Bowser-Chao$^{(1)}$,
Darwin Chang$^{(2,3)}$,
and
Wai-Yee Keung$^{(1)}$ }
\begin{center}
\it
$^{(1)}$Physics Department, University of Illinois at Chicago,
IL 60607-7059, USA\\
$^{(2)}$Center for Theoretical Sciences and Physics Department,
N. Tsing-Hua University, Hsinchu 30043, Taiwan, R.O.C.\\
$^{(3)}$Institute of Physics, Academia Sinica, Taipei, R.O.C.\\
\vspace{.5cm}
\end{center}

\begin{abstract}
We discuss the subtle interplay between  strong and weak CP phases
that has often been ignored in the literature.  We also point out the
potentially important role that it plays in various models.
\end{abstract}

\vspace{1in}
\centerline{Festschrift for T.-Y. Wu}
\centerline{Published in {\it Chinese J. Phys. {\bf 35}, 842 (1997)}}

\newpage

\section*{Introduction}


\quad
\vskip -1cm

The mystery of CP violation has only grown with time.  It was pointed
out by Landau\cite{r:landau} in 1957 that  the issue of CP violation is
tightly coupled with that of  complex phases in field theory.  The
latter, in turn, has proven to be one of the most subtle aspects of the
theory --- in part, due to its connection with  anomalies.  In strong
interactions,
it is particularly complicated  because the relation between the phases
of the field operators and physical states is highly non-trivial, stemming from
the non-perturbative effects of confinement and chiral condensation.
This is in particular true for chiral phases --- in sharp contrast to
vector phases, whose associated vectorial symmetries (e.g., flavor
symmetries) are {\em not} dynamically broken by QCD. Chiral phase
rotations can generate rather convoluted effects in general  hadronic
matrix elements, a fact which leads to a very subtle interplay between
the strong and weak CP phases, as recently pointed out by the
authors\cite{bck1}.

In this article, we wish to illustrate this result, using as an example a
recently proposed superweak model of CP violation\cite{bck2}, and then
discuss its significance for other models, including the standard
Kobayashi-Maskawa  Model\cite{km}.

\section*{The Illustrative Model}

Consider the following superweak model of CP violation\cite{bck2} with a
$U(1)_X$ family gauge symmetry.  The quark fields transform as
\begin{eqnarray}
&X = +1: s_R, b_R \quad\quad &X = -1: c_R, t_R \\
&X = +2: u_R      \quad\quad &X = -2: d_R
\end{eqnarray}
A Higgs doublet $\phi_1$ transforming as  $X=-1$ is needed to give mass
to the $s$, $b$,  $c$, and $t$ quarks.  Another Higgs doublet $\phi_2$
with $X=2$ is responsible for the  $d$ and $u$ quark masses.  We shall
not worry about the lepton sector (see Ref.\cite{bck2} for discussion of
such issues.)  The charges are arranged so that the anomaly is
cancelled\cite{geng}.  In particular, the Tr$(XXX)$ anomaly is cancelled
between the up and down quarks. We also assume that CP symmetry is imposed
on the Lagrangian  so that, before breaking of the $U(1)_X$ symmetry,
all couplings are real, and the strong CP-violation parameter
$\theta_{QCD}$ is identically zero.

An $SU(2)\times U(1)$ singlet Higgs is introduced to break the $U(1)_X$
symmetry at a high energy scale of  around 1-10 TeV.  The $X$ boson
generates flavor changing neutral currents and gives rise to the
superweak interaction\cite{bck2} of the form
$$
{\cal L}^{\Delta S=2}_{\rm sw}
= \frac{F_R}{M_X^2} (\bar{s_R} \gamma_\mu d_R)^2\  + \ {\rm h.c.}
$$
where $F_R$ is a dimensionless coupling constant.   There are also
superweak interaction involving right-handed up type quarks, but which
are not relevant here.  Due to the rich phenomenology in the kaon
system, this particular superweak interaction is the most relevant one
experimentally. The $U(1)_X$ symmetry breaking may or may not break CP,
and so the coupling $F_R$ may or may not be complex\cite{bck2}.  We consider
both possibilities below.

For our purposes, we can focus on the reduced effective theory, with
CP-conserving Standard Model-type interactions and vanishing
$\theta_{QCD}$, to which has been added the new superweak interaction
${\cal L}^{\Delta S=2}_{\rm sw}$. We shall consider the scenario in which
the up quark is massive, unless otherwise specified, while, for the sake of
our argument, we will consider the limit in which $m_d$ is zero.
Without the new interaction the parameter $\theta_{QCD}$ is then
unphysical, with CP a good symmetry.
By the usual argument, with a
massless quark present --- in this case, the $d$ quark --- the
right-handed component of that quark can be rotated to absorb
$\theta_{QCD}$ via the axial anomaly, while otherwise leaving the
lagrangian invariant. With the addition of the interaction ${\cal
L}^{\Delta S=2}_{\rm sw}$, however, $\theta_{QCD}$ becomes  physical, as can
be seen by considering the following cases:

(a) When $F_R$ is real and $\theta_{QCD}=0$, CP is an unbroken symmetry
of nature.  If we rotate $d_R$ by an arbitrary phase, $F_R$ becomes
complex and $\theta_{QCD}$   nonzero, but all other couplings are
unchanged. Consider a ``calculation'' of $\epsilon$ using the usual
techniques for the hadronic matrix elements, and with the lagrangian
with $d_R$ rotated as described. This naive procedure will yield a
non-zero $\epsilon$ due to the non-vanishing contribution of ${\cal
L}^{\Delta S=2}_{sw}$ to  Im$(M_{12})$ of the $K-\bar{K}$ mixing matrix
--- a result which, of course, is incorrect. CP is conserved,
appearances notwithstanding. {\em Something} must occur, then, to
correct the  ``usual techniques for the hadronic matrix elements''. New
complex phases must enter into the calculation of the hadronic matrix
elements to cancel the CP-odd contribution due to the unphysical
complexity in
$F_R$.  The lesson here is that when one makes a chiral rotation of
quarks, not only does $\theta_{QCD}$ change, but the hadronic matrix
elements must also undergo certain phase changes. If the strong force were
to conserve CP, with
breaking due only to small, weak (non-QCD) corrections, one could certainly
choose a basis of states such that hadronic matrix elements are purely real
(modulo absorptive contributions). In the case considered above,
however, QCD (with $\theta_{QCD}$ non-zero )
does not respect CP, via instanton contributions. The extent to which
matrix elements must
be adjusted in the presence of a non-zero $\theta_{QCD}$ is precisely
that which cancels out the spurious non-zero $\epsilon$ in the example
above.
(One could alternatively reabsorb $\theta_{QCD}$ by a rotation of
$u_R$, which would generate a complex up quark mass. Since we have
assumed $m_d=0$, the imaginary part of the mass matrix cannot be
proportional to the identity matrix.
Vacuum stability\cite{theta} then
requires {\em reinterpretation} of the usual meson states such that complex
hadronic matrix elements are explicitly required.)

(b) If $F_R$ is complex and $\theta_{QCD}=0$, CP is violated. In this
case, the correct (non-zero) value for $\epsilon$ can be calculated
without complication;  all hadronic matrix elements (modulo absorptive
contributions) can correctly be assumed to be real.  It is illuminating
to consider calculation of $\epsilon$ in another basis, which is obtained
by a phase rotation of $d_R$ such that $F_R$ becomes real and $\theta_{QCD}$
non-zero.  Since the two theories are the same, one must arrive at the
same result for $\epsilon$.  One can thus draw
a rather surprising conclusion: $\theta_{QCD}$
 {\em can also, in certain situations, contribute to $\epsilon$}.

In fact, from the way we obtain the $\theta_{QCD}$ contribution to
$\epsilon$ in this example, one realizes that there is an important
subtlety here. {\it The actual contribution from $\theta_{QCD}$ to
$\epsilon$ is correlated to the explicit mechanism of CP violation},
which in our current example is the superweak $F_R$.
A related result is that  when $\theta_{QCD}$ is not zero, how each hadronic
matrix element develops a phase {\em also depends on the particular electroweak
mechanism of CP violation in the theory}.  In the present case, the CP
violating coupling also happens to be the chiral symmetry
breaking phase.

Another lesson one learns is that the usual argument (see for example
Ref.\cite{donoghue}), which concludes that the contribution of
$\theta_{QCD}$ to CP-violating quantities such as the neutron electric
dipole moment (edm) must
be proportional to $m_u m_d$, is not strictly correct, a counterexample
towhich is offered by the simplified model presented above.  The role of
$m_d$ is
replaced by the  coupling $F_R$. Of course, $F_R$
breaks the chiral symmetry associated with $d$ quark, so that the $d$
quark will certainly pick up mass at some (probably higher-loop) level, but
the point is that the $F_R$ coupling plays a much more direct role in the
contribution of
$\theta_{QCD}$ than even the induced $m_d$!

Now we come to an apparent paradox whose resolution gives even further
insight into the interplay between strong and weak CP phases.

We parenthetically noted above that if redefinition of the quark phases
generates an imaginary part of the quark mass matrix {\em not proportional}
to the identity matrix, the low-energy meson states must be suitably
reinterpreted to ensure stability of the vacuum around which we carry out
perturbation theory. This redefinition explicitly reintroduces the phase(s)
rotated from the couplings into certain hadronic matrix elements, to ensure
rephasing invariance. If both $m_d, m_u$ vanish, then arbitrary rotation of
the corresponding right-handed quarks seems to have no effect on vacuum
stability, since the mass matrix is left real and diagonal (only $m_s$
non-zero). Then, apparently, all phases may be arbitrarily rotated away,
and with them, any possibility of CP violation. Specifically, consider the
following variant of the two cases already considered:

(c) Let $F_R$ be complex, but take $\theta_{QCD}$ to be zero. If {\em both}
$m_u, m_d$ are strictly zero, is CP conserved or violated?
At first glance, one might  claim the phase in $F_R$ to be unphysical, since a
combined phase rotation of the form $u_R\rightarrow e^{-i\delta}u_R$ and
$d_R\rightarrow e^{i\delta}d_R$ can make $F_R$ real and
maintain $\theta_{QCD}=0$.  It is very tempting to claim  that CP
violation is proportional to $m_u$  for a small up quark mass and further
that there is no CP violation when $m_u\rightarrow 0$ because the phase of
$F_R$ then becomes.

This conclusion is {\em incorrect}, however,  because we have ignored the
vacuum
degeneracy in the case of massless $u$ and $d$ quarks. Different
choices of vacua would give different CP violation.  It is true that
there exists one very special vacuum where CP is conserved. However, a
general vacuum posseses chiral condensate with a phase uncorrelated to
that of $F_R$, and thus CP violation usually occurs, {\em even if} $F_R$ is
real (since what is important is the relative phase between the vacuum and
$F_R$). This idea can be
demonstrated directly in the chiral effective lagrangian approach.
The chiral field $\Sigma$ (3$\times$3 unitary matrix) can be perturbed
around a vacuum configuration diag$(e^{-i\phi},e^{i\phi},1)$. If
$F_R$ is turned off, the strong interaction is independent of $\phi$
because of the chiral symmetry.  However, with $F_R$, the phase $\phi$
has physical meaning and has implications with respect to CP violation.
Now we include effects of the real quark masses $m_u\ne 0$ and $m_d\ne 0$.
Their net effect is simply to pick out a particular vacuum,
with   $\Sigma=$diag$(1,1,1)$. In this case of vacuum alignment, a complex
$F_R$ is necessary, but also sufficient, for CP violation, since again it
is the relative phase between $F_R$ and the vacuum that is important. In
some sense, the (possibly infinitesimal) up and down quark masses {\em
enforce} CP violation, in the particular case that $F_R$ is real, whereas
in the massless case, CP violation is still generally expected via the
vacuum phase.

\section*{Discussions of other models}

The above argument can not only  be applied to the superweak $F_R$ interaction
above, it can also be easily generalized to a wide variety of models  ---
especially those in
which $d_R$ has additional interactions, including lepto-quark models,
charged Higgs models, etc.  It is the interplay between the low energy
hadronic physics and CP violation that is the issue  here and that is
also why it is so easy to get confused.

Consider the KM model\cite{km}.  CP is explicitly broken by the
dimension-four Yukawa couplings and
thus the $\theta_{QCD}$ is present and
uncalculable.  {\it A priori}, based on the previous argument, one does
not know at all what kind of phases to expect for each hadronic matrix
elements.  The usual way to proceed is to make assumptions.  After the
quark mass matrices are diagonalized, and after making several chiral
phase rotations to turn the charge current mixing matrix into the
Kobayashi-Maskawa form, with only one complex phase (the
Kobayashi-Maskawa phase) remaining, one then {\it assumes} that
$\theta_{QCD}$ is made vanishingly small order by order in this very
special basis.  With the lagrangian in this form, QCD conserves CP, so that
a basis for strong eigenstates may be chosen in which all hadronic matrix
elements can be made real. Note,
however, that the change of phase convention from one convention (e.g.,
that with the  KM phase), to another (e.g, the Chau-Keung\cite{chaukeung}
convention) does
not involve any chiral phase
rotations.

Another example of the impact of our argument is to consider the
Weinberg- Branco Three Doublet Model\cite{weinberg-branco} of
spontaneous CP violation.  This and many other examples are treated in
Ref.\cite{bck2}.

To summarize, in some models of CP violation, the CP violating mechanism
itself also breaks chiral symmetry.  In such cases, there can be very
interesting and subtle interplay between the strong and weak CP violating
effects that deserve careful study.

D.~B.-C. and W.-Y.~K. are supported by a grant from the Department of
Energy, and D.~C. by a grant from the National Science Council of R.O.C.
We wish to thank Bill Bardeen and Tom Imbo for several insightful
discussions, and Lincoln Wolfenstein,  Glenn Boyd, Peter Cho,  Ben Grinstein,
Jim Hughes, and Rabi Mohapatra for helpful conversations.  D.~C. also wishes to
thank the High Energy Group and Physics Group of Argonne National Lab.
for hospitality while this work was initiated and the National Center
for Theoretical Scienecs at Hsinchu, R.O.C. for partial support.


\def\baselinestretch{1.2}

\end{document}